\documentclass[%
reprint,
superscriptaddress,
amsmath,amssymb,
aps,
prd,
]{revtex4-1}

\usepackage{graphicx}% Include figure files
\usepackage{dcolumn}% Align table columns on decimal point
\usepackage{comment}
\usepackage{enumitem}
\usepackage{bm}% bold math
\usepackage{siunitx}
\usepackage{xcolor}

\newcommand{\db}[1]{ \textcolor{red}{(DB: #1)} }
\newcommand{\gk}[1]{ \textcolor{blue}{(GK: #1)} }

\begin{document}

\title{Reply to Robinson and Michaud, arXiv:2002.08893}

\author{Noah~Kurinsky}\thanks{kurinsky@fnal.gov}
\affiliation{Fermi National Accelerator Laboratory, Batavia, Illinois 60510, USA}
\affiliation{Kavli Institute for Cosmological Physics, University of Chicago, Chicago, Illinois 60637, USA}

\author{Daniel~Baxter}\thanks{dbaxter@kicp.uchicago.edu}
\affiliation{Kavli Institute for Cosmological Physics, University of Chicago, Chicago, Illinois 60637, USA}
\affiliation{Enrico Fermi Institute, University of Chicago, Chicago, Illinois 60637, USA}

\author{Yonatan~Kahn}\thanks{yfkahn@illinois.edu}
\affiliation{University of Illinois at Urbana-Champaign, Urbana, Illinois 61801, USA}

\author{Gordan~Krnjaic}\thanks{krnjaicg@fnal.gov}
\affiliation{Fermi National Accelerator Laboratory, Batavia, Illinois 60510, USA}
\affiliation{Kavli Institute for Cosmological Physics, University of Chicago, Chicago, Illinois 60637, USA}

\author{Peter~Abbamonte}\thanks{abbamont@illinois.edu}
\affiliation{University of Illinois at Urbana-Champaign, Urbana, Illinois 61801, USA}

\date{\today}

\begin{abstract}
We respond to Robinson and Michaud's (RM) comment (arXiv:2002.08893) on our recent preprint arXiv:2002.06937, in which we discuss recent excesses in low-threshold dark matter searches, and offer a potential unifying dark matter interpretation. We thank RM for their feedback, which highlights the critical need for future measurements to directly calibrate plasmon charge yields for low $\sim$ 10 eV energy depositions. RM objected to our assertion that plasmons generated at energy scales below 100~eV may have a large branching fraction into phonons. As we argue below, the points raised by RM do not invalidate our primary conclusions, as they pertain to a much different energy scale than we discuss in our paper.  
\end{abstract}

\maketitle

\section{Introduction}

We would like to acknowledge the feedback on our preprint \citet{kurinsky2020dark} which we recently received from \citet{robinson2020comment} (hereafter RM), concerning the charge yield model through the plasmons which we postulate are excited by some unknown signal source. We are grateful for this feedback, which we believe reflects the importance of our observations for the dark matter (DM) detection community, and highlights the uncertainties inherent in our models in this unusual kinematic regime.

In this Reply, we first summarize the key points of our preprint relevant to the criticisms of RM, provide a point-by-point response, and conclude by reiterating our points of agreement and disagreement.

\section{Summary}
\label{sec:Summary}

Our preprint \citet{kurinsky2020dark} makes four key observations:

\begin{enumerate}[label=(\alph*)]
    \item   There are several low-energy excesses in semiconductor experiments, with rates that are consistent to within a factor of two independent of detector environment, shielding, or overburden;
    
    \item  Reconciling the charge and heat spectra from the two runs of the EDELWEISS germanium detector requires a charge yield model which is strongly inconsistent with either electron recoil (ER) or elastic nuclear recoil (NR); 
    
    \item The spectral shape of the excess in total detector energy $E_{det}$ resembles the tail of the plasmon, a known resonance in condensed matter systems, so it is possible the source of the excess is plasmon excitation;
    
    \item  There are two plausible DM scenarios which can match the observed total rate in germanium, where the spectrum is primarily determined by the plasmon lineshape and not by the DM velocity distribution:
        \begin{itemize}
            \item{\bf DM Scenario 1:} DM scatters a {\it nucleus} whose primary low-energy recoil induces {\it secondary} plasmon excitation with associated phonon production. As we argue below, {\it none of the claims in RM apply to this scenario in any way.}
            
            \item{\bf DM Scenario 2:} Direct plasmon excitation through a fast millicharged DM particle in analogy with plasmon excitation in electron energy loss spectroscopy. As we argue below, while superficially similar to the scenarios considered in RM, no simultaneous measurement of energy loss and ionization yield is inconsistent with this scenario.
        \end{itemize}

\end{enumerate}
The criticisms of RM exclusively concern observations (c) and (d); we will take as given that our observations (a) and (b) stand.

The crux of RM's argument is the claim that energetic electrons passing through semiconductors primarily lose energy through plasmon excitation, depositing energy in multiples of the plasmon energy $E_p$. The plasmons then subsequently decay into electron-hole pairs, such that the average energy deposited per electron-hole pair created is 3 eV in germanium. In other words, the measured yield curve for electrons in germanium is claimed as evidence that plasmons created by fast electrons decay into an average of 5 electron-hole pairs per plasmon, which is inconsistent with our yield model which reconciles the EDELWEISS spectra and requires an average of 0.25 electron-hole pairs per plasmon (assuming Poisson statistics), or exactly 1 electron-hole pair in every plasmon event.

There are two exceptions we take to this claim. The first is that, for high-energy charged particles, hard scattering (direct scattering with electrons) and bremsstrahlung are much more important than collective effects for determining energy loss, especially for minimum ionizing particles (MIPs) (see chapter 33 of Ref.~\cite{PhysRevD.98.030001}). While plasmons may play a small role in energy loss, the primary loss is due to direct ionization of electrons and defect creation. The second objection is that these interactions result in small momentum losses for the MIP, but large momentum transfers to the condensed matter system (compared to the natural momentum scale set by the inverse lattice spacing). Even if plasmons are produced, they would typically be created with large momenta, off-resonance, and quickly decay by Landau damping into electron-hole pairs. In contrast, the plasmons we invoke in our paper carry very little momentum, and in a perfect crystal (using the random phase approximation \cite{PhysRev.111.442} to model plasmons, as is customary) would have an infinite lifetime. This suggests the dominant decay path of a low-momentum plasmon would not be through creation of electron-hole pairs, and could instead be dominated by anharmonic decay into phonons.

%\textbf{NK: This paragraph needs to be changed; we disagree that this is the correct picture, and we already rule out MIPs based on existing data.}
%RM are correct that this observation disfavors our Scenario 2, where DM is millicharged and excites plasmons exactly as would a fast electron, but with a much longer inelastic mean free path. However, our Scenario 1 concerns a completely different kinematic regime, where a hard DM scatter on a nucleus creates a secondary plasmon excitation in association with multiple phonons, much like the Migdal effect converts nuclear recoil to charge excitation with very different kinematics compared to direct ER \cite{Baxter:2019pnz,Essig:2019xkx}. There is no experimental validation of a NR yield curve below 100 eV, and thus it is plausible that the plasmon branching fraction to phonons could depend strongly on these kinematics, which are maximally distinct from the kinematics of fast ionizing electrons or x-rays. Indeed, a main point of our preprint is to highlight the need for NR calibrations in the energy regime below 100 eV, and we reiterate this point here. Furthermore, RM's observations about Scenario 2 suggest that exclusion limits on DM-electron scattering are much stronger than previously claimed, precisely because many models of millicharged DM have a high-velocity component which could excite plasmons directly, inconsistent with the the lack of a corresponding charge excess in the silicon detectors.

While a better understanding of decay paths of plasmons in real materials is needed, there are good reasons to conclude that the objections of RM are not applicable to the phenomena described in either of our DM models.

\section{Response to specific claims in 2002.08893}

\begin{enumerate}
\item \textbf{Claim:} ``Plasmons are a coherent excitation between electrons and ions that have been well studied in electron transmission and inelastic x-ray scattering physics. They are expected to be seen in the spectra of eV-sensitive calorimeters exposed to keV and MeV-energy photons.''
\textbf{Response:} RM are certainly correct that plasmons have been observed in all these channels. However, to our knowledge, the indirect excitation of plasmons through a hard nuclear scattering (our scenario 1 from observation (d) above) has not yet been observed. In addition, the plasmon down-conversion mechanism for electron-recoils is a convenient heuristic without any predictive power (see e.g. \cite{rothwarf1973plasmon}). In the energy range of interest for low-threshold detectors, the charge yield for $\sim$ 15 eV energy depositions near the plasmon energy, especially for those with very low momentum transfer, is thus far unvalidated. 

\item \textbf{Claim:} ``By choosing an ionization yield of 0.25 electron-hole pairs per 16 eV plasmon generated, they are able to support their dark matter interpretation.''
\textbf{Response:} We emphasize that our yield model to reconcile the two EDELWEISS spectra is independent of any DM interpretation. We demonstrated in \cite{kurinsky2020dark} that, assuming a common source, the two spectra are \emph{completely inconsistent} under the hypotheses of either elastic ER or NR. Thus, some other inelastic process must be active; given that the branching fraction of secondary plasmons created from NR below 100 eV is thus far unconstrained, and that the spectral shape of the excess resembles the high-energy tail of the plasmon, it is not unreasonable to suppose that the plasmon may play some role. The plasmon is not a fundamental particle, so its decays need not obey Lorentz invariance; the plasmon branching ratio to phonons and/or charge may be a strong function of the energy and momentum of the plasmon, and of the interactions with the additional phonons created in the hard scattering event in Scenario 1.

\item \textbf{Claim:} ``As seen in electron energy loss spectroscopy, the dominant energy loss mechanism for ionizing electrons is plasmon excitations. Thus, plasmon ionization is merely an intermediate step of electron ionization, and their ionization yields must nearly identical [\textit{sic}].'' \textbf{Response:} Ref.~\cite{kundmann1988study}, cited by RM, does not support their claim. Ref.~\cite{kundmann1988study} is a study of plasmon lineshapes in electron energy loss spectroscopy for semi-relativistic electrons of energy $20-100$ keV. The measurements of Ref.~\cite{kundmann1988study} are simply energy loss measurements, and do \emph{not} determine whether the plasmon subsequently decays into electron/hole pairs and/or phonons. 

%However, given that the probability of plasmon creation is hundreds of times higher than the probability of single electron/hole excitation \cite{pines1956collective}, and that a simple model for plasmon decay to electron/hole pairs matches the observed ionization yield \cite{rothwarf1973plasmon}, it is reasonable to surmise that the plasmon branching fraction to electron/hole pairs is order-1, disfavoring our Scenario 2.

\item \textbf{Claim:} ``Additionally, regardless of the proportion of energy deposited in plasmons and other electronic excitations, similar ionization yields should result.'' \textbf{Response:} Again, Ref.~\cite{ABS} cited by RM does not support this claim. Ref.~\cite{ABS} is a phenomenological model for electron ionization where plasmons (necessarily with high momentum) immediately decay to a single electron-hole pair and there is only a single phonon energy. There is no reason to believe this model accurately captures the features of real semiconductors, especially in the low-energy regime where no calibration data exists and the density of states of phonons is large.

\item \textbf{Claim:} ``With a large electron yield per plasmon, it is difficult to interpret any of the observed excess signals as dark matter in light of the strong constraint from DAMIC for excesses of multi-electron/hole events.'' \textbf{Response:} As we only construct a yield model for EDELWEISS, our yield model is only for germanium, and says nothing about the corresponding yield in silicon, which may be markedly different. Furthermore, in our paper \cite{kurinsky2020dark} we address the potential inconsistency with DAMIC and note that the analysis procedure, rather than the yield model, could relieve some of the tension with the multi-electron/hole rates of the other silicon experiments. 

%Once again, a low charge yield in DAMIC \cite{PhysRevLett.123.181802} (and CDMS HVeV \cite{Agnese_2018} and SENSEI \cite{Abramoff_2019}) would disfavor our Scenario 2 with direct plasmon excitation, as one would expect a strong peak in the 5-electron bin from 15 eV plasmons decaying with $\sim$ 3 eV per electron-hole pair. As we have emphasized, this data is not in conflict with our DM models, where no calibration data exists.

\item \textbf{Claim:} ``Any excitation in this high-momentum high-energy regime should be expected to produce single electron states. Even if the phonon plus plasmon final state were probable, the phonon's momentum would lie well outside the first Brillouin zone.'' \textbf{Response:} Here it is worth making more precise the setup of our Scenario 1. Suppose DM with incoming kinetic energy of (say) 60 eV undergoes a hard scattering with a nucleus. There is sufficient kinetic energy to remove a nucleus from the lattice, but (a) if the DM is light, the energy transfer to the nucleus is highly inefficient, and (b) the presence of the plasmon resonance means that it is energetically favorable to deposit only 15 eV to excite a plasmon at low momentum, and let the nucleus absorb the momentum transfer. Since the nucleus is still trapped at its lattice site, it must shed the momentum in the form of \emph{a large number of phonons,} not a single phonon. What we are proposing is an inherently nonlinear, multi-body process. Single electron states are certainly kinematically accessible, but the strong plasmon resonance in the dielectric function means that plasmon excitation is \emph{more} likely than single particle/hole excitation, when kinematically permitted. A full treatment of this process must also take into account the finite size of the Brillouin zone through Umklapp processes, which violate crystal momentum conservation by integer multiples of reciprocal lattice vectors, when accounting for momentum transfer to a detector crystal.

\item \textbf{Claim:} ``The momentum required to extract 16 eV from dark matter\dots where 2.27 eV/$c$ is the inverse lattice spacing of silicon.'' \textbf{Response:} We assume this was a typographical error; the inverse lattice spacing of silicon is closer to 2.27 keV/$c$. Indeed, we are aware of these kinematics as we note in Footnote 7 of \cite{kurinsky2020dark}.

\item \textbf{Claim:} ``The results of Kurinsky et al should not be taken as evidence for dark matter, although it does highlight the ongoing need to investigate the effect of collective modes how we detect radiation [\textit{sic}].'' \textbf{Response:} Nowhere in our preprint \cite{kurinsky2020dark} do we claim evidence for DM. We emphasize once again the motivations for our work: (a) to point out an intriguing coincidence of excesses across several experiments, (b) to propose a condensed matter phenomenon which may explain the spectral shape of the excesses, and (c) to construct toy models of DM which may explain the observed rates. We wholeheartedly agree with RM on the need to investigate the effect of collective modes, but it is equally important to note that the effects we are proposing, especially in the kinematic regime of Scenario 1, are plausible from the perspective of condensed matter and deserve dedicated calibration experiments. The source of these persistent excesses remains mysterious, and to date, our plasmon model is the only one which attempts to quantitatively reconcile a number of unusual features of these excesses, \emph{independent of any DM interpretation.}

\end{enumerate}

\section{Conclusion}

Finally, we would like to reiterate what statements we do and do not make in our preprint based on the perceptions noted in RM:
\begin{enumerate}
    \item We do not claim that the excesses we note are definitively sourced by DM, and we certainly do not claim a discovery. We point out that many different experiments see strikingly similar rates, and show that it is not inconsistent with a well-motivated model of DM. The differing overburden, technology, and shielding between even the charge readout experiments invoke some background which is unaffected by operating environment, crystal history, operating temperature, location, or other known experimental conditions. This in itself is a significant and interesting observation.
    \item Our observation (a) in Sec.~\ref{sec:Summary} stands independently of our yield interpretation. However, we show that the observed spectra are inconsistent with ER, and in strong tension with any elastic NR interpretation. In order to reconcile liquid noble experiments with cryogenic and CCD experiments, we rely on the feature that is most discrepant between them: the existence of a strong resonance in solids.  Even if the plasmon decay were to produce more electron-hole pairs than are observed, that does not rule out some alternative inelastic, resonant condensed matter effect with a low yield.  It is significant that the integrated rates at high voltage and zero volts from EDELWEISS are compatible, and thus it is valid to constrain the effective yield based on that observation. It is also significant that zero-yield models are disfavored, which rule out events like crystal cracking or vibration-induced phonons.
    \item It is also possible that some other source of events is generating these excesses, but we have argued in Ref.~\cite{kurinsky2020dark} that any known SM interactions are strongly disfavored. If some other inelastic interaction is occurring due to SM particles, this is still representative of interesting new physics related to low-energy particle interactions which, as we have emphasized here and in our preprint, would significantly enhance the sensitivity of existing detectors to light DM.
    %\item We agree with RM that the DAMIC result is in tension with the other silicon charge experiments. We expect that further results from SENSEI, SuperCDMS, and DAMIC-M will be able to explain why such an interesting rate coincidence ($\sim7$ Hz/g in DAMIC, $\sim7-10$ Hz/g above single electron leakage for SuperCDMS and SENSEI) is observed. The DAMIC result is by far the most shielded, studied system of the three, and thus it would be reasonable to expect that if SuperCDMS and SENSEI are seeing a combination of the same background with elevated single electron or $\gamma$-induced leakage, then further study with enhanced shielding might reduce the single electron rates to bring them into closer agreement with the DAMIC result. Given that the inherent leakage due to thermal promotion of carriers is expected to be 4 orders of magnitude lower than either SENSEI or DAMIC \cite{SENSEILDM,janesick2001scientific}, and that this model cannot account for charge in the two-electron bin, we find simple leakage strongly discrepant with the observed rates and charge production.
\end{enumerate}

%Other comments received on our prepreint, such as Ref~\cite{SunnyBlog}, are in line with our interpretation of the importance of this work.

\begin{acknowledgments}
We would like to thank Alan Robinson and \'Emile Michaud for discussing their Comment with us, and for early responses to our inquiries clarifying their objections to our original preprint. 
%We would like to thank Alan Robinson for discussing his Comment with us, and for early responses to our inquiries clarifying his objections to our original preprint.
While none of the conclusions in our preprint have changed as a result of this Comment, we believe we can clarify points of confusion to better explain the mechanism we propose.
\end{acknowledgments}

\bibliography{main}

\end{document}